\documentclass[%
reprint,
superscriptaddress,
showkeys,
 amsmath,amssymb,
 aps,
]{revtex4-2}

\usepackage{graphicx}
\usepackage{color}
\usepackage{dcolumn}
\usepackage{longtable}
\usepackage{bm}
\usepackage{hyperref}
\usepackage{physics}
\hypersetup{
    colorlinks=true,
    linkcolor=blue,
    citecolor=blue
    }
\newcommand{\comment}[1]{}

\begin{document}

\preprint{APS/123-QED}

\title{Thermal and atomic effects on coupled-channels heavy-ion fusion}

\author{Iain Lee}
\affiliation{%
Department of Physics, University of Surrey, Guildford, GU2 7XH, UK
}%

\author{Gilbert Gosselin}%
\affiliation{%
CEA, DAM, DIF, F-91297, Arpajon, France
}%

\author{Alexis Diaz-Torres}
\affiliation{%
Department of Physics, University of Surrey, Guildford, GU2 7XH, UK
}%

\date{\today}

\begin{abstract}
Stellar nuclear fusion reactions take place in a hot, dense plasma within stars. To account for the effect of these environments, the theory of open quantum systems is used to conduct pioneering studies of thermal and atomic effects on fusion probability at a broad range of temperatures and densities. Since low-lying excited states are more likely to be populated at stellar temperatures and increase nuclear plasma interaction rates, a $^{188}\text{Os}$ nucleus was used as a target that interacts with an inert $^{16}\text{O}$ projectile. Key results showed thermal effects yield an average increase in fusion probability of 15.5\% and 36.9\% for our test nuclei at temperatures of $0.1$ and $0.5$ MeV respectively, compared to calculations at zero temperature. Thermal effects could be tested in a laboratory using targets prepared in excited states as envisaged in facilities exploiting laser-nucleus interactions.
\end{abstract}


\maketitle



\textit{Introduction}. Understanding the cosmic origins of heavy elements is one of the biggest problems in science. Nuclear fusion reactions, which often occur in high energy density plasmas (HEDPs), are some of the most relevant reactions contributing to nucleosynthesis and stellar evolution. These environments are hot, and contain a mixture of ions and electrons. Low-energy fusion experiments in laboratories are often subject to the target medium, and the current focus is on the use and improvement of radioactive beams to produce the nuclei needed for these reactions \cite{SINGH201799,Zagrebaev_2007}. Recreating the stellar environments where these reactions take place is challenging, and hence computational models offer a path to understanding this complex problem. Our unique study showcases a novel method to include the influence of an external physical environment (a plasma) on low-energy fusion reactions, using the theory of open quantum systems. This study suggests that coupling-assisted tunneling in nuclear fusion is strongly enhanced by thermal effects. These effects could be observed in experiments with targets prepared in excited states, e.g., exploiting laser-nucleus interactions \cite{Buervenich2006}.


The impact of plasma on a reaction may be dominated by either thermal or Coulomb effects. A first order estimation of their relative importance can be carried out by calculating a Coulomb parameter, $\Lambda$, which is essentially a ratio between Coulomb and thermal energies \cite{Salpeter1954,Ichimaru_1982,Whitten_1984}, 

\begin{equation}
    \Lambda \equiv \frac{\langle Z_{i}e \rangle ^{2}}{a_{i}\mathcal{T}}, 
\end{equation}
where $ \langle Z_{i}e \rangle$ is the average ion charge in the plasma, $a_{i}$ is the average inter-ionic distance and $\mathcal{T}$ is the temperature in MeV. When $\Lambda \ll 1$, the Coulomb energy is insignificant to thermal energy and a Debye-H\"{u}ckel potential is assumed \cite{Rolfs2006-kt}, and for $\Lambda \gg 1$, the Coulomb energy dominates the plasma interaction, and this regime is modeled with an ion-sphere potential. Electron screening potentials effectively reduce the Coulomb barrier, and these become increasingly complex in HEDPs. An extensive inclusion of electron screening would include higher density effects such as ion-ion and ion-electron correlations \cite{Jancovici_1977}, and relativistic effects such as pair production at high temperatures ($\approx 1 \text{ MeV}$ or higher). Reviews on weak- and strong-screening regimes in HEDPs can be found in Refs. \cite{Aliotta22,RMP2011,Itoh79}. For the $^{16}\text{O}$ projectile and $^{188}\text{Os}$ target used in this work, the screening effects may be significant at some temperatures ($0  - 1$ MeV) and densities ($10 - 10^{5} \text{ gcm}^{-3}$) studied. For simplicity, we focus our study on the effects of plasma temperature and the role of nuclear plasma interactions (NPIs) on low-energy fusion reactions, with the latter expected to affect stellar nucleosynthesis \cite{NEET,Adriana_2014}. Examples of processes involving NPIs are nuclear excitation by electron capture (NEEC) or transition (NEET). These have been observed experimentally but still are not well-understood \cite{PhysRevA.73.012715,neet1}. The $^{16}$O+$^{188}$Os reaction is used as a test case because (i) this reaction simplifies the model calculations, and (ii) the $^{188}$Os target allows one to maximise both NPI and thermal effects.

The thermal population of low-lying excited states has been previously considered in neutron capture studies using the Hauser-Feshbach statistical model \cite{PhysRevC.82.015803,PhysRevC.82.015804}. The neutron cross sections were weighted with temperature-dependent population probabilities of target's excited states, leading to a stellar enhancement factor. However, these effects were ignored in heavy ion fusion studies, since the reactions of interest would typically involve inert nuclei or nuclei with high excitation energy states of several MeV. 

For the first time, the present work studies thermal and NPI effects on nuclear fusion using a dynamical, quantum coupled-channels model. The coupled-channels density-matrix method has demonstrated the ability to calculate energy-resolved fusion probabilities using an open quantum system approach \cite{LEE2022136970}. 





\textit{Thermal effects.} A thermal environment is expected to change the initial population of the intrinsic energy eigenvalues, $\{e_\alpha\}$, of nuclei that it encompasses. To model the thermal effects on a fusion reaction, we introduce Boltzmann factors, $w_{\alpha}$, into the initial density matrix, 

\begin{equation} \label{Eq:initialdm}
    \rho ^{rs}_{\alpha\alpha} (t=0) = w_\alpha\ket{r} \bra{s},
\end{equation}
where $\ket{r}$ refers to a Gaussian or Coulomb wave packet describing the internuclear motion on a radial grid \cite{LEE2022136970}. We have tested two different initial wave packets with the same Gaussian envelope but two different boosts, $\mathcal{B}(k_{0}r)$, namely a plane wave ($e^{-ik_{0}r}$) and an incoming Coulomb wave, $H^{-}_{L=0} (k_{0}r)$: 

\begin{equation}
    \psi (r,r_{0},\sigma _{0},k_{0}) = \mathcal{N}^{-1} \, \exp [- \frac{(r-r_{0})^2}{2 \sigma _{0}^{2}}] \mathcal{B}(k_{0}r),
\label{eq2}    
\end{equation}
where $\mathcal{N}$ is a normalisation constant, $r_0$ is the initial, central position of the wave packet, $r$ is a radial grid position, $\sigma _0$ is the spatial dispersion, and $k_{0}$ is the average wave number, which depends on the average incident energy $E_0$, $r_0$ and $\sigma_0$ and is found by solving $E_0 = \bra{\psi}\hat{H}\ket{\psi}$, $\hat{H}$ being the Hamiltonian of the collision. Eq. (\ref{eq2}) with a plane-wave boost is a Gaussian wave packet, while this is a Coulomb wave packet when a Coulomb-wave boost is used.

In Eq. (\ref{Eq:initialdm}), the density matrix is diagonal in the energy eigenstate basis, denoted by $\ket{\alpha}$. The initial population of the energy eigenstates is given as,

\comment{
\begin{equation}
    w_{i} = \dfrac{\exp{-\dfrac{e_{i}}{k_{b}t}}}{\text{Tr}\left[\exp{-\dfrac{H_{0}}{k_{b}t}}\right]} ,
\end{equation}
}

\begin{equation} \label{eq:t_weights}
     w_{\alpha} = \dfrac{ (2 I_{\alpha} + 1) \, \exp{-\cfrac{e_{\alpha}}{\mathcal{T}}}}{\displaystyle\sum _{\alpha^{'} = 1}^{N} (2 I_{\alpha^{'}} + 1) \, \exp{-\cfrac{e _{\alpha^{'}}}{\mathcal{T}}}},
\end{equation}
where $I_{\alpha}$ is the spin value. Since the present calculations only include excited states of the $^{188}$Os ground-state rotational band, the spin degeneracy factors in Eq. (\ref{eq:t_weights}) will be irrelevant. Excited states are thermally populated before the target and projectile interact with each other and thermodynamic equilibrium is assumed at the start of the reaction \cite{Dzhioev_2016,Dzhioev_2020}. As known from coupled channels calculations, coupling of the radial motion to energy eigenstates can cause changes in fusion probability. Population of excited states in either the target or projectile nucleus due to surface vibrations or rotational excited states lead to an overall increase in fusion probability due to coherent coupled channels effects \cite{RevModPhys.70.77}. For the dynamical calculations, we use the same equation of motion for the density matrix of the reduced system as the one used in Ref. \cite{LEE2022136970}, Eq. ($5$). This method uses the Lindblad master equation, allowing coupling between the radial and energy eigenstate bases, and uses Lindblad operators to introduce novel effects into the calculation. An energy projection technique is used to calculate the fusion probability for specific collision energies \cite{LEE2022136970}.

In the following calculations, we use a $^{188}\text{Os}$ target nucleus due to its low-lying $2^{+}$ rotational excited state at $155$ keV \cite{THOMPSON1975444} and a $^{16}\text{O}$ projectile nucleus due to its high $6.13$ MeV first excited state. Hence we only consider the ground state of $^{16}\text{O}$ and the ground and first excited state of $^{188}\text{Os}$. The parameters used in these calculations are the same as in Ref. \cite{LEE2022136970}, except from the potential parameters which are unique to the projectile and target pair. The Woods-Saxon nuclear interaction potential parameters used in this work are: $\text{V}_{0}=60.64 \text{ MeV},\text{R}_{0}=1.2\text{ fm},\text{a}_{0}=0.63\text{ fm}$, and these parameters provide the same height of the uncoupled Coulomb barrier ($\text{V}_{B}=71.7$ MeV) as the microscopic S\~{a}o Paulo potential \cite{PhysRevC.66.014610}. The deformation parameter of the $155$ keV excited state is $\beta _{2} = 0.184$ \cite{national_nuclear_data_center}.


\textit{Atomic effects}. NPI effects are included in the calculations by introducing new matrix elements into Eq. ($5$) of Ref. \cite{LEE2022136970},

\begin{align}   
    \Gamma ^{rr}_{12} &= \gamma _{12}  \label{env_int1} \\
    \Gamma ^{rr}_{21} &= \gamma _{21} ,
    \label{env_int2}
\end{align}
where $\gamma _{12}$ and $\gamma _{21}$ are the respective excitation and de-excitation rates between the ground state and first excited state of $^{188}\text{Os}$. These affect the excited state population, dependent on the temperature and density of the plasma. The relevant rates for this work are shown in Table \ref{tab:Coulomb_screening}, and were calculated using the ISOMEX code \cite{Gosselin_2004}, which is based on the relativistic average atom model and assumes local thermal equilibrium. The considered excitation processes due to NPIs are: resonant photon absorption, inelastic electron scattering, NEEC and NEET. The de-excitation processes include spontaneous photon emission, induced photon emission, internal conversion, bound internal conversion (BIC), and super-elastic electron scattering. Since the excited state of $^{188}$Os is much higher than the binding energy of the K-shell atomic orbitals, NEET and BIC do not contribute to the plasma induced nuclear transition rates.

\begin{table*}[htb!]
\caption{Excitation ($\gamma _{12}$, left column) and de-excitation ($\gamma _{21}$, right column) rates for NPIs between the ground state and first excited state of a $^{188}\text{Os}$ nucleus for different temperatures and densities, calculated using the ISOMEX code \cite{Gosselin_2004}.}
\label{tab:Coulomb_screening}
\resizebox{0.9\textwidth}{!}{%
\centering
\begin{ruledtabular}
\def\arraystretch{1.5}
\begin{tabular}{cccc}

Density & $\mathcal{T} = 0.01$ MeV                         & $\mathcal{T} = 0.1$  MeV                        & $\mathcal{T} = 1$ MeV                           \\ \hline 
10 g/cm$^3$                      & 6.0 $\cdot$ $10^{2}$ / 6.5 $\cdot$ $10^{8} s^{-1}$ & 7.3 $\cdot$ $10^{8}$ / 6.9 $\cdot$ $10^{8} s^{-1}$ & 1.6 $\cdot$ $10^{10}$ / 3.8 $\cdot$ $10^{9} s^{-1}$  \\ \hline 
$10^{3}$ g/cm$^3$                     & 7.0 $\cdot$ $10^{2}$ / 7.6 $\cdot$ $10^{8} s^{-1}$ & 7.3 $\cdot$ $10^{8}$ / 6.9 $\cdot$ $10^{8} s^{-1}$ & 1.6 $\cdot$ $10^{10}$ / 3.8 $\cdot$ $10^{9} s^{-1}$ \\ \hline
$10^{5}$ g/cm$^3$                    & 8.0 $\cdot$ $10^{2}$ / 8.6 $\cdot$ $10^{8} s^{-1}$ & 7.8 $\cdot$ $10^{8}$ / 7.3 $\cdot$ $10^{8} s^{-1}$ & 1.6 $\cdot$ $10^{10}$ / 3.8 $\cdot$ $10^{9} s^{-1}$ \\ 

\end{tabular}%
\end{ruledtabular}
}
\end{table*}

It was found that the effects of NPIs were negligible ($< 10^{-6}$ \% increase in fusion probability) for a $^{16}\text{O}$ projectile and $^{188}\text{Os}$ target, at all temperatures and densities considered. Considering that the timescale of the fusion reactions is of the order of $10^{-22}$ s, the effective excitation rates are too low to have an impact on the population of the excited state \cite{Gosselin_2007} and therefore the overall effect on fusion is weak. 

Nuclear fusion reactions are commonly initiated from the ground state of the collision partners. Should scenarios exist where fusion reactions are initiated from long-lived intrinsic high angular momentum excited states, the NPIs would be more effective \cite{Chiara2018}. Additionally, a recent experiment \cite{PhysRevLett.128.212502} showed that missing excited electronic configurations could be a reason for discrepancy between theory and experiment for NEEC reactions. 

\begin{figure}[htb!]
 \centering
 \includegraphics[width=0.50\textwidth]{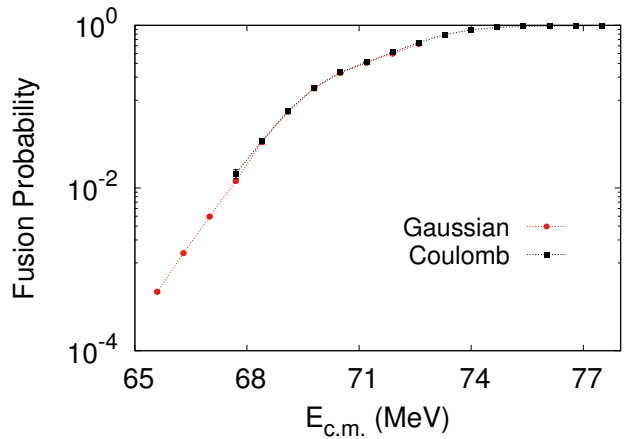}
 \caption{\label{fig:baseline}A construction of the energy-resolved fusion probability for a $^{16}\text{O}$ projectile and $^{188}\text{Os}$ target with coupled channels but without a plasma environment ($\mathcal{T} = 0$ MeV). Gaussian and Coulomb wave packets are used, taking the average energy-resolved fusion probability for a range of incident mean energies, $E_0$. Error bars due to statistical error are included but most are insignificant. The nominal Coulomb barrier between these nuclei is $71.7$ MeV.}
\end{figure}

\textit{Results and discussion}. We construct the coupled channels fusion probability of a zero temperature, environment-less fusion reaction for an inert $^{16}\text{O}$ projectile and $^{188}\text{Os}$ target with two states (ground and first excited state), given in Fig. \ref{fig:baseline}. Multiple wave packets with different initial mean energies ($E_{0}$) were used to check that the results converge \cite{LEE2022136970}. This serves as verification of the method and a baseline that allows us to evaluate the thermal effects of the plasma.

\begin{figure}[ht!]
\centering
\includegraphics[width=0.35\textwidth]{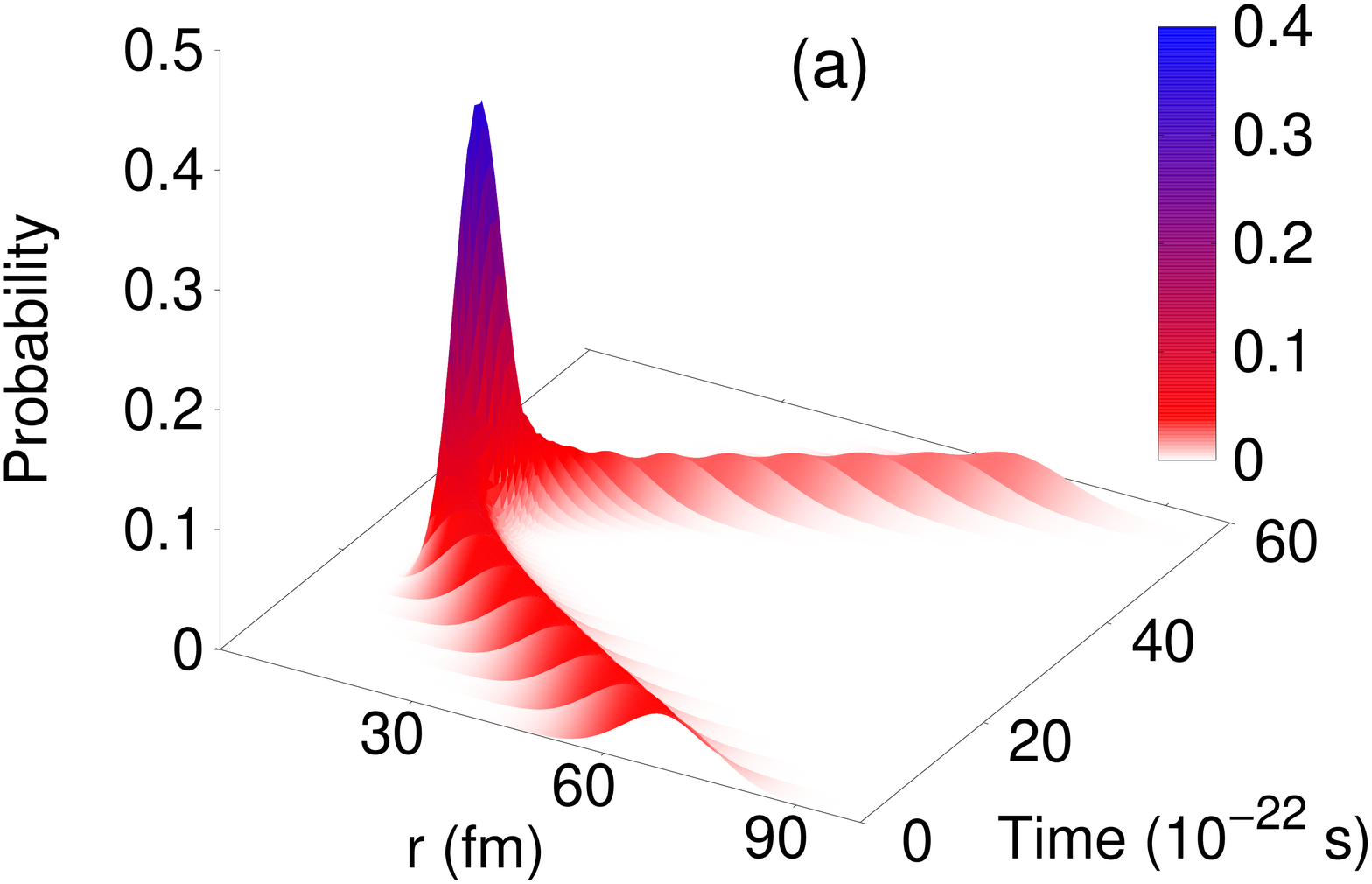} \\
\includegraphics[width=0.35\textwidth]{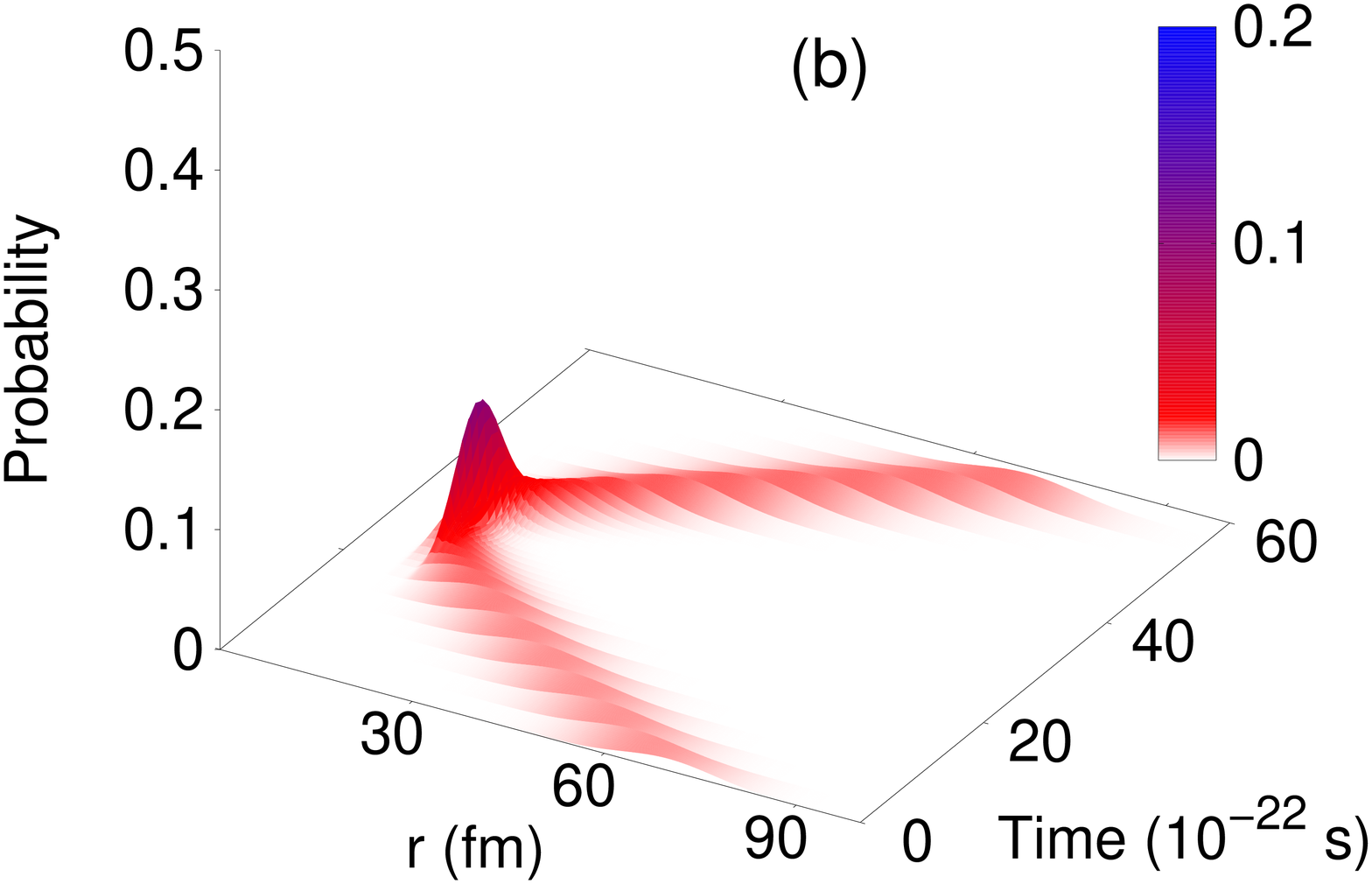}
\caption{Radial position probability as a function of internuclear radius and time for a head-on collision of $^{16}$O + $^{188}$Os with $E_0=70$ MeV and $\mathcal{T} = 0.1$ MeV. The radial probability changes for (a) the elastic and (b) inelastic channels respectively, as the nuclei approach their Coulomb barrier ($r \approx 11.5$ fm). The inelastic channel is thermally populated before the target and projectile interact with each other. For visualisation, when the mean radius is larger than $20$ fm, the time step is $3 \times 10^{-22}$ s.}
\label{fig:propag}
\end{figure}

Fig. \ref{fig:propag} shows the population of the radial grid basis states over time for an initial Gaussian wave packet with an average energy $E_0=70$ MeV and a temperature $\mathcal{T} = 0.1$ MeV. The change of population of both the ground state (elastic channel) and the $2^{+}$ excited state of $^{188}$Os (inelastic channel) due to both thermal effects and the radial coupling between these states can be observed.

\begin{figure}[htb!]
 \centering
 \includegraphics[width=0.50\textwidth]{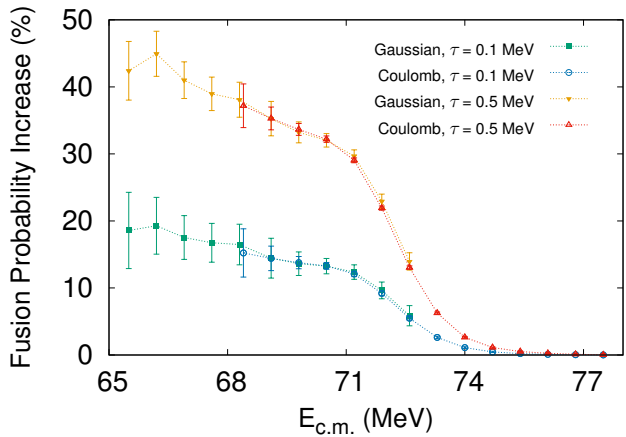}
 \caption{\label{fig:fusion_diff}The increase in fusion probability for a $^{16}\text{O}$ projectile and $^{188}\text{Os}$ target due to the presence of a thermal plasma environment. For each temperature ($\mathcal{T} = 0.1$ MeV and $0.5$ MeV), the fusion probability increase is calculated using the ratio of averaged thermal environment calculations to averaged baseline calculations with no environment (Fig. \ref{fig:baseline}). For the Gaussian wave packet, calculations were initiated with $E_{0}=$ 60, 63, 65, 67 and 70 MeV and for the Coulomb wavepacket, $E_{0}=$ 65, 67 and 70 MeV.}
\end{figure}

To isolate the effects of temperature on fusion probability, the increase in energy-resolved fusion probability was calculated using the ratio between coupled channels calculations at either $\mathcal{T} = 0.1$ MeV or $\mathcal{T} = 0.5$ MeV and $\mathcal{T} = 0$ MeV, shown in Fig. \ref{fig:fusion_diff}. The fusion probability was calculated by taking the average energy-resolved fusion probability for initial wave packets with varying $E_{0}$. For $\mathcal{T} = 0.1$ MeV, the green (square) points were calculated using a Gaussian wave packet with $E_{0}=$ 60, 63, 65, 67 and 70 MeV, and the blue (circle) points were calculated using a Coulomb wave packet with $E_{0}=$ 65, 67 and 70 MeV. The same method was used for both $\mathcal{T} = 0$ and $\mathcal{T} = 0.5$ MeV. The error bars associated with these points are simply due to the standard error in the mean. We use a Gaussian wave packet for its accuracy at deep sub-barrier energies compared to a Coulomb wave packet. However, a Coulomb wave packet offers a better global description of fusion probability around and above the Coulomb barrier, as discussed in Ref. \cite{LEE2022136970}. Below the Coulomb barrier, the average increase in fusion probability was 15.5\% and 36.9\% for the $0.1$ MeV and $0.5$ MeV temperatures respectively. Above the Coulomb barrier, the increase quickly diminishes to a few percent. 

\begin{figure}[htb!]
 \centering
 \includegraphics[width=0.50\textwidth]{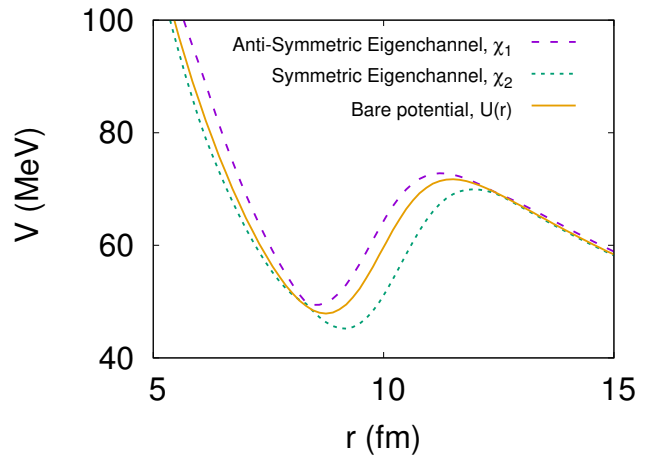}
 \caption{\label{fig:barriers} The Coulomb barriers of the decoupled eigenchannels. The symmetric barrier is the lowest, dominating the fusion process at energies below the nominal, 71.7 MeV Coulomb barrier of the bare potential, U(r).}
\end{figure}

The results in Fig. \ref{fig:fusion_diff} are an advancement on work that showed that sub-barrier fusion is enhanced when channel couplings are included, due to the fusion contribution of excited states \cite{DASSO1983381}. The thermal increase in fusion probability can be explained by studying the radial wave function of the entrance channel,

\comment{
\begin{equation}
    \lambda _{1,2} (r)
\end{equation}}

\begin{equation} \label{eq:entrance_channel}
    \Psi _{0}(r) = \sqrt{1- w_{2}} \cdot \psi _{1} (r) + \sqrt{w_{2}} \cdot \psi _{2} (r) ,
\end{equation}
where $\psi _{1}$ and $\psi _{2}$ are the radial wave functions of the ground state and excited state, and these wave functions also contain their respective energy basis states, $\ket{1}$ and $\ket{2}$. The excited state Boltzmann factor calculated in Eq. (\ref{eq:t_weights}) is denoted by $w_{2}$. Two effective Coulomb barriers are created from a linear combination of two dynamically coupled wave functions, $\psi _{1}$ and $\psi _{2}$, as shown in Fig. \ref{fig:barriers}. This combination is anti-symmetric and symmetric, defining two decoupled eigenchannels, $\chi _{1,2} (r)$,

\begin{equation} \label{eq:chi}
    \chi _{1,2} (r) = \frac{1}{\sqrt{2}} \left( \psi _{1} (r) \mp  \psi _{2} (r) \right).
\end{equation}

The height of the Coulomb barrier for the symmetric eigenchannel ($\chi_2$) is significantly smaller than that of the anti-symmetric eigenchannel ($\chi_1$), leading to an increase of the fusion probability relative to a single channel calculation involving the state $\psi _{1}(r)$ only. 

Temperature affects the fraction of the eigenchannels contained in the entrance channel configuration in Eq. (\ref{eq:entrance_channel}), 
   
\begin{equation}
    |\bra{\chi_{1}}\ket{\Psi_{0}}| ^{2} = \frac{1}{2} - \sqrt{(1-w_{2}) \, w_{2}} \, ,
\end{equation}

\begin{equation}
    |\bra{\chi_{2}}\ket{\Psi_{0}}| ^{2} = \frac{1}{2} + \sqrt{(1-w_{2}) \, w_{2}} \, .
\end{equation}

The inclusion of temperature leads to an increase in the initial population of the excited state, $w_2$, and consequently there is a larger fraction of the eigenchannel $\chi_2$ (with the lowest Coulomb barrier) in the entrance channel configuration. The effect is more prominent with increasing temperature, which is supported by our results in Fig. \ref{fig:fusion_diff}. Hence there is an enhancement of fusion probability in comparison to coupled channels calculations without temperature ($w_2 = 0$).


\textit{Summary}. We have addressed an area of unexplored territory by assessing environmental quantum effects on heavy-ion fusion calculations. The coupled-channels density-matrix method, which is based on the theory of open quantum systems, unambiguously include thermal and atomic effects on subbarrier fusion dynamics. The calculations show that plasma temperature strongly enhances fusion probability. This pioneering effort suggests that careful considerations should be made when modeling or performing experiments on collision partners with initially populated low-lying excited states (e.g., see Ref. \cite{Buervenich2006}). Despite no changes in fusion probability due to atomic effects, there is scope to reintroduce these effects when more experiments are completed and new theoretical studies are published.

\textit{Acknowledgment}. IL and AD-T acknowledge support from the Leverhulme Trust (UK) under Grant No. RPG-2019-325.

\bibliography{references}

\end{document}